\documentclass[aps,pra,twocolumn,superscriptaddress]{revtex4}

\usepackage{epsfig}
\usepackage{amsmath,amssymb,amsfonts}

\newcommand{\beq}{\begin{equation}}
\newcommand{\eeq}{\end{equation}}

\newcommand{\ket}[1]{\left|{#1}\right\rangle}

\newcommand{\tr}[1]{{\rm tr}\left(#1\right)}

\begin{document}

\title{Experimental proposal for symmetric minimal two-qubit state tomography}
\author{Amir Kalev}
\affiliation{Centre for Quantum Technologies, National University of Singapore, 3 Science Drive 2, 117543, Singapore}
\author{Jiangwei Shang}
\affiliation{Centre for Quantum Technologies, National University of Singapore, 3 Science Drive 2, 117543, Singapore}
\author{Berthold-Georg Englert}
\affiliation{Centre for Quantum Technologies, National University of Singapore, 3 Science Drive 2, 117543, Singapore}
\affiliation{Department of Physics, National University of Singapore, 2 Science Drive 3, 117542, Singapore}
\date{Finalized on Dec. 13, 11; Posted on Mar. 08, 12}
\begin{abstract}
We propose an experiment that realizes a symmetric informationally complete (SIC) probability-operator measurement (POM) in the four-dimensional Hilbert space of a qubit pair. The qubit pair is carried by a single photon as a polarization qubit and a path qubit. The implementation of the SIC POM is accomplished with the means of linear optics. The experimental scheme exploits a new approach to SIC POMs that uses a two-step process: a measurement with full-rank outcomes, followed by a projective measurement on a basis that is chosen in accordance with the result of the first measurement. The basis of the first measurement and the four bases of the second measurements are pairwise unbiased --- a hint at a possibly profound link between SIC POMs and mutually unbiased bases.
\end{abstract}
\maketitle

Quantum state tomography,  the procedure for inferring the state of a quantum system from measurements applied to it, is an important component in most, if not all, quantum computation and quantum communication tasks.  The successful execution of such tasks hinges in part on the ability to  assess with high efficiency the state of the system at various stages. 

A general measurement in quantum mechanics is a probability-operator measurement (POM). A POM is informationally complete (IC) if any state of the system is determined completely by the probabilities for the POM outcomes \cite{ic1,ic2,ic3}. State tomography infers these probabilities from the data acquired with the aid of the POM. 

A symmetric IC POM (SIC POM) is an IC POM of a particular kind. In a $d$-dimensional Hilbert space (of kets) it is composed of $d^2$ subnormalized rank-1 projectors, $\{{\cal P}_j\}_{j=1}^{d^2}$, with equal pairwise fidelity of $1/(d+1)$. Their high symmetry and high tomographic efficiency have attracted the attention of many researchers, and a lot of work, both analytical and numerical, has been devoted to the construction of SIC POMs  in various dimensions, see e.g. \cite{berge04,renes04,appleby05,scott06,scott10,zauner11}. 

In contrast to the  major theoretical progress, up to date,  all experiments and even proposals for experiments implementing  SIC POMs have been limited to the very basic quantum system, the two-level system (qubit) \cite{ling06,pimenta10}, with the exception of \cite{steinberg11} where a SIC POM for a three-level system was approximated. This is, in part, due to the fact that there is no systematic procedure for implementing SIC POMs in higher dimensions, in a simple experimental set-up. 

In this contribution we suggest a feasible experiment that implements a SIC POM for four-dimensional Hilbert space of a qubit pair. The qubit pair is carried by a single photon, and the measurement is realized by passive linear optical elements and photodetectors. The experiment is clearly feasible with current technology.

The proposal exploits a new theoretical approach for constructing SIC POMs \cite{our}. In this approach, we `break' the SIC POM into two successive measurements, each with $d$ outcomes, with the intention that each measurement would be relatively easy to implement. Unexpectedly, we find that the two successive measurements that make up the SIC POM in dimension 4, not only have a strikingly simple form, but also hint at a close relation between the structure of mutually unbiased bases (MUB) and that of the SIC POM; Ref.~\cite{durt10} is a recent review on MUB.

Let us begin by setting up nomenclature  and notations. A general measurement on a quantum system is composed of a set of outcomes. The latter are mathematically represented by positive operators ${\cal P}_j$ that sum up to the identity operator.  The probability of obtaining the outcome ${\cal P}_j$ is given by the Born rule:  \mbox{$p_j = \tr{{\cal P}_j\rho}$}, where $\rho$ is the pre-measurement statistical operator of the system. If the $j$th outcome is found, the post-measurement statistical operator of the system is given by
\mbox{$\rho_{j}=\frac{1}{p_j}P_j\rho P^\dagger_j$}, 
where $P_j$ is the relevant Kraus operator for the $j$th outcome, \mbox{${\cal P}_j=P^\dagger_j P _j$}. Note that the decomposition of the ${\cal P}$s into the corresponding Kraus operators is not unique; for example, \mbox{$P^\dagger_j P_j$} is invariant under the unitary transformation \mbox{$P_j\rightarrow U_jP_j$}, with different $U_j$s corresponding to different implementations of the POM.

Suppose that a given system is subjected to a sequence of two POMs, each with $d$ outcomes,  \mbox{$\{{\cal A}_k=A^\dagger_k A_k\}_{k=1}^{d}$}, followed by \mbox{$\{{\cal B}^{\scriptscriptstyle(k)}_j\}_{j=1}^{d}$}, where the superscript $k$ indicates that in general the second measurement depends on the actual outcome of the first measurement. Following Born's rule, the probability of obtaining the $n$th and $m$th outcomes for the first and second measurements is given by \mbox{$\tr{\rho A^\dagger_n {\cal B}^{\scriptscriptstyle(n)}_mA_n}$}. Accordingly, the two successive measurements are equivalent to a single POM with $d^2$ outcomes \mbox{${\cal P}_{n,m}=A^\dagger_n {\cal B}^{\scriptscriptstyle(n)}_mA_n$} with \mbox{$n,m=1,\;\ldots,\;d$}. Indeed, upon finding the over-all outcome ${\cal P}_{n,m}$, we know that the $n$th outcome of the first POM and the $m$th outcome of the second POM are the case. In what follows we will identify the $A$s and the ${\cal B}$s such that the ${\cal P}$s make up the SIC POM for a qubit pair. But before doing so, we discuss such an identification for a single qubit.

\begin{figure}[t]
\centering
\includegraphics[scale=1.]{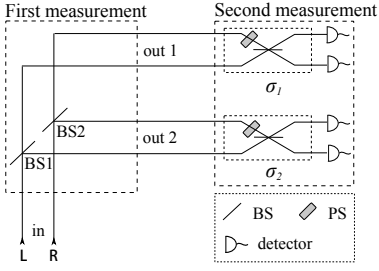}
\caption{An optical implementation of the tetrahedron measurement using two successive measurements.}
\label{fig:tetrahedron}
\end{figure}

All SIC POMs in two-dimensional Hilbert space are unitarily equivalent to the ``tetrahedron measurement'' (TM), whose outcomes correspond to four vectors that define a tetrahedron in the Bloch sphere \cite{berge04,renes04}. The TM could be realized by a sequence of two measurements, as  sketched in Fig.~\ref{fig:tetrahedron}. Here, the qubit is encoded in a spatial alternative of a single photon (``path qubit''): traveling on the left or on the right. A unitary transformation on the qubit state amounts to sending the photon through a set of beam splitters (BSs) and phase shifters (PSs) \cite{englert01}. 

In this optical setting the TM is implemented as follows: First, two BSs (BS1 and BS2) are used to implement the Kraus operators  \mbox{$A_1={\rm diag}(t_1,t_2)$} and \mbox{$A_2={\rm diag}(r_1,r_2)$}, where ${\rm diag}$ stands for a diagonal matrix, and $t_i$ and $r_i$ are the transmission and reflection amplitudes of the $i$th BS. A photon which enters the apparatus with a path statistical operator $\rho$, exits at port $k$  with the statistical operator \mbox{$A_k\rho A_k^\dagger/\tr{A_k\rho A_k^\dagger}$}.  For the values  \mbox{${\textstyle t_1{=}r_2{=}\sqrt{\frac1{2}-\frac1{\sqrt{12}}}}$} and \mbox{${\textstyle t_2{=}r_1{=}\sqrt{\frac1{2}+\frac1{\sqrt{12}}}}$}, these operators correspond to the measurement outcomes \mbox{${\cal A}_k{=}\frac1{2}(1+\frac1{\sqrt{3}}(-1)^k\sigma_3)$} with \mbox{$k{=}1,2$} (the $\sigma$s are the Pauli operators with  $\sigma_3$ diagonal in the left-right basis). Then a photon that exits the first measurement apparatus at port 1 is measured in the $\sigma_1$ basis while a photon  that exits at port 2 is a measured in the $\sigma_2$ basis. These measurements could be realized by balanced BSs and appropriate PSs, as indicated in the figure. 

Actually, the TM was successfully implemented in an optical system \cite{ling06}, where the qubit was encoded in a photon's polarization (``polarization qubit'') rather than in spatial alternatives. The set-up of \cite{ling06} also consisted of a sequence of two measurements, quite analogous to what is described above. In that set-up, a partially polarizing beam splitter (PPBS) was used to implement the Kraus operators $A_k$ and then, depending on whether the photon was transmitted or reflected, a measurement of  $\sigma_1$ or $\sigma_2$ followed. 

Before moving on to the qubit-pair case, let us close the present discussion with three remarks: (i) In the above construction, the qubit MUB play a central role; they are used to construct, by means of successive measurements, the SIC POM. We will see below that such a relation appears in dimension 4 as well. (ii) A practical implementation of the scheme presented in Fig.~\ref{fig:tetrahedron} requires the stabilization of  the interferometer loop defined by the four BSs. (iii) SIC POMs for a three-level system could be implemented by using a similar set-up, but with allowing the photon to take three different paths \cite{our}.

\begin{figure*}[t]
\centering
\includegraphics[scale=1.]{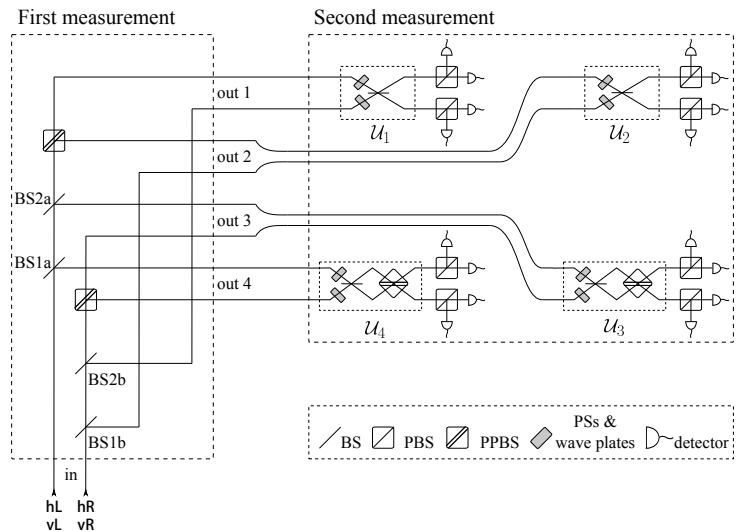}
\caption{A successive-measurement scheme for realizing the SIC POM of a qubit pair. Here the two-qubit state is encoded in the spatial-polarization state of a single photon.}
\label{fig:sicpom_dim4}
\end{figure*}

In dimension 4, there is only one known SIC POM, and all the other known SIC POMs are unitarily equivalent to it \cite{appleby05}. This SIC POM is composed of 16 subnormalized projectors onto 16 (fiducial) kets. The latter are represented in the following matrices  as columns  with \mbox{$N=\sqrt{5+\sqrt{5}}$} and {$\chi=\sqrt{2+\sqrt{5}}$} \cite{bengtsson10},%
\begin{alignat}{2}\label{fiddim4}
&\frac1{N}{\left(%
\begin{array}{rrrr}
\chi&\chi&\chi&\chi\\
1&-1&1&-1\\
1&1&-1&-1\\
1&-1&-1&1
\end{array}%
\right)}\!,
&\;&\frac1{N}{\left(%
\begin{array}{rrrr}
1&1&1&1 \\
1&-1&1&-1\\
i\chi&i\chi&-i\chi&-i\chi\\
-i&i&i&-i
\end{array}%
\right)}\!,\nonumber\\
&\frac1{N}{\left(%
\begin{array}{rrrr}
1&1&1&1 \\
i\chi&-i\chi&i\chi&-i\chi\\
i&i&-i&-i\\
-1&1&1&-1
\end{array}%
\right)}\!,
&\;&\frac1{N}{\left(%
\begin{array}{rrrr}
1&1&1&1 \\
i&-i&i&-i\\
1&1&-1&-1\\
-i\chi&i\chi&i\chi&-i\chi
\end{array}%
\right)}\!.
\end{alignat}%
Each of these matrices could be written as a diagonal matrix times a unitary matrix. The set of bases, corresponding to each unitary matrix, together with the computational basis, form the complete set of MUB in dimension 4. To be more specific, the diagonal matrices are
\begin{align}\label{Adim4}
A_1=&\frac1{N}{\rm diag}(\chi,1,1,1),\;\;A_3=\frac1{N}{\rm diag}(1,1,\chi,1),\nonumber\\
A_2=&\frac1{N}{\rm diag}(1,\chi,1,1),\;\;A_4=\frac1{N}{\rm diag}(1,1,1,\chi),
\end{align}
and the  unitary matrices are
\begin{alignat}{2}\label{Bdim4}
{\cal U}_1\!&=\!\frac1{2}{\left(%
\begin{array}{rrrr}
1&1&1&1 \\
1&-1&1&-1\\
1&1&-1&-1\\
1&-1&-1&1
\end{array}%
\right)}\!,\;
&{\cal U}_3\!&=\!\frac1{2}{\left(%
\begin{array}{rrrr}
1&1&1&1 \\
1&-1&1&-1\\
i&i&-i&-i\\
-i&i&i&-i
\end{array}%
\right)}\!,\nonumber\\
{\cal U}_2\!&=\!\frac1{2}{\left(%
\begin{array}{rrrr}
1&1&1&1 \\
i&-i&i&-i\\
i&i&-i&-i\\
-1&1&1&-1
\end{array}%
\right)}\!,\;
&{\cal U}_4\!&=\!\frac1{2}{\left(%
\begin{array}{rrrr}
1&1&1&1 \\
i&-i&i&-i\\
1&1&-1&-1\\
-i&i&i&-i
\end{array}%
\right)}\!.
\end{alignat}%
Noting that \mbox{$\sum_jA^\dagger_jA_j=1$}, we identify the $A$s with the Kraus operators of a measurement.

Actually, the operations of Eq.~\eqref{Bdim4} transform the computational basis into the MUB, 
\begin{align}\label{MUBdim4}
{\mathfrak B}_1&=\left\{\begin{array}{r}
\frac1{\sqrt{2}}(\ket{0}+\ket{1})\\
\frac1{\sqrt{2}}(\ket{0}-\ket{1})\\
\end{array}\right\}\otimes\left\{\begin{array}{r}
\frac1{\sqrt{2}}(\ket{0}+\ket{1})\\
\frac1{\sqrt{2}}(\ket{0}-\ket{1})\\
\end{array}\right\},\nonumber\\
{\mathfrak B}_2&=\left\{\begin{array}{r}
\frac1{\sqrt{2}}(\ket{0}+i\ket{1})\\
\frac1{\sqrt{2}}(\ket{0}-i\ket{1})\\
\end{array}\right\}\otimes\left\{\begin{array}{r}
\frac1{\sqrt{2}}(\ket{0}+i\ket{1})\\
\frac1{\sqrt{2}}(\ket{0}-i\ket{1})\\
\end{array}\right\},\nonumber\\
{\mathfrak B}_3&={\rm CZ}\left\{\begin{array}{r}
\frac1{\sqrt{2}}(\ket{0}+i\ket{1})\\
\frac1{\sqrt{2}}(\ket{0}-i\ket{1})\\
\end{array}\right\}\otimes\left\{\begin{array}{r}
\frac1{\sqrt{2}}(\ket{0}+\ket{1})\\
\frac1{\sqrt{2}}(\ket{0}-\ket{1})\\
\end{array}\right\},\nonumber\\
{\mathfrak B}_4&={\rm CZ}\left\{\begin{array}{r}
\frac1{\sqrt{2}}(\ket{0}+\ket{1})\\
\frac1{\sqrt{2}}(\ket{0}-\ket{1})\\
\end{array}\right\}\otimes\left\{\begin{array}{r}
\frac1{\sqrt{2}}(\ket{0}+i\ket{1})\\
\frac1{\sqrt{2}}(\ket{0}-i\ket{1})\\
\end{array}\right\},
\end{align}
where CZ stands for the controlled-Z (phase flip) operation, \mbox{${\rm CZ}={\rm diag}(1,1,1,-1)$}. The  bases ${\mathfrak B}_1$ and ${\mathfrak B}_2$ are composed of product states, while the bases ${\mathfrak B}_3$ and ${\mathfrak B}_4$ consist of maximally entangled states.

The structure of fiducial vectors  in Eq.~\eqref{fiddim4} (which form the SIC POM in dimension 4) allows us to implement the SIC POM by two successive measurements: A measurement whose Kraus operators are given in Eq.~\eqref{Adim4}, and depending on the measurement outcome, a measurement in one of the MUB of Eq.~\eqref{MUBdim4}. Next, we propose an optical implementation for this scheme.

Our proposal is based on the methods of \cite{englert01} where the two qubits, a polarization qubit and a path qubit, are encoded in a single photon. (We chose here to use a polarization qubit instead of  another path qubit in order to avoid as many interferometric loops as possible in the optical set-up.) We consider the vertical (${\sf v}$) and horizontal (${\sf h}$) polarizations as the basic alternative of the polarization qubit, and  traveling on the left (${\sf L}$) or on the right (${\sf R}$) as the basic  alternative of the path qubit. A unitary transformation on the two-qubit state amounts to sending the photon through a set of passive linear optical elements that unitarily change the state of the path and polarization qubits \cite{englert01}. For our purpose we need the following optical elements:
half-wave plates (HWPs), BSs, polarization and path dependent PSs, and PPBSs. A PPBS is a BS whose reflection and transmission coefficients depend on the polarization. Its action corresponds to a joint unitary transformation on the polarization-path qubits. In the present context, it suffices to consider a PPBS with real amplitude division coefficients $r$ and $t$ that obey the unitarity condition \mbox{$r^2+t^2=1$} for the vertical and horizontal polarizations,
\beq\label{ppbs} 
U_{\!{\textrm{\tiny PPBS}}}=\left(\begin{array}{cccc}
r_v&t_v&0&0\\
-t_v&r_v&0&0\\
0&0&r_h&t_h\\
0&0&t_h&-r_h\\
\end{array}%
\right).
\eeq
This is a block-diagonal matrix, with the blocks transforming the vertical or horizontal polarization, respectively. Two  cases of interest are (i) \mbox{$r_v=t_h=1$}: the polarizing beam splitter (PBS)  which totally reflects (transmits) vertically (horizontally) polarized light, and (ii) \mbox{$r_v=r_h=1$}: the CZ gate.

The Kraus operators for the first measurement are listed in Eq.~\eqref{Adim4}. Their realization  is schematically drawn in Fig.~\ref{fig:sicpom_dim4} at the `first measurement' part. For each port, we set the parameters of the different optical elements such that a photon which enters the apparatus with a polarization-path statistical operator $\rho$, exits at port $k$  with the two-qubit statistical operator \mbox{$A_k\rho A_k^\dagger/\tr{A_k\rho A_k^\dagger}$}. To be more specific, the apparatus is configured such that the beam splitters BS1a and BS1b have the same properties and so have beam splitters BS2a and BS2b. The PPBSs on the left and right arms also have the same properties. The reflection coefficient of BS1a  and BS1b is \mbox{$r_1=1/N$}. The reflection coefficient $r_2$ of BS2a  and BS2b satisfies \mbox{$t_1r_2=1/N$}, that is, \mbox{$r_2=1/\sqrt{N^2-1}$}, where $t_1$ is the transmission coefficient of BS1a(b). Setting \mbox{$r_v=t_h=y$} in Eq.~\eqref{ppbs}, the two  PPBSs  transform vertically polarized incident light $\ket{{\sf v}}$ to the polarizations \mbox{$y\ket{{\sf v}}$} and \mbox{$\sqrt{1-y^2}\ket{{\sf v}}$} in the reflected and transmitted arms, respectively, and horizontally polarized light $\ket{{\sf h}}$ to the polarizations \mbox{$\sqrt{1-y^2}\ket{{\sf h}}$} in reflection and \mbox{$y\ket{{\sf h}}$} in  transmission. The amplitude division coefficient $y$ is chosen such that \mbox{$t_1t_2y=1/N$}, and therefore, \mbox{$y=1/\sqrt{N^2-2}$}, where $t_2$ is the transmission coefficient of BS2a(b).  These settings ensure that the measurement of Eq.~\eqref{Adim4} is realized.

To complete the measurement scheme, a second measurement is taking place. This measurement depends on the actual outcome of the first measurement, namely, on the output port where the photon exits. For photons emerging from the $k$th port, basis ${\mathfrak B}_k$ of Eq.~\eqref{MUBdim4} is measured. In order to measure in a given basis, ${\mathfrak B}_k$, we first apply a unitary operation ${\cal U}_k$ of Eq.~\eqref{Bdim4} that transforms the basis ${\mathfrak B}_k$ into the computational basis and then measure in the computational basis by using PBSs and photodetectors, as illustrated in Fig.~\ref{fig:sicpom_dim4} at the `second measurement' part.

To implement the unitary transformations of Eq.~\eqref{Bdim4}, one could use either a single, specially designed, birefringent material, or a sequence of wave plates and PPBSs. Considering the latter option, these unitary transformations are
\begin{align}\label{unitaries}
{\cal U}_1&=U_{\!{\textrm{\tiny HWP}}}\otimes U_{\!{\textrm{\tiny BS}}},\nonumber\\  
{\cal U}_2&=\left(U_{\!{\textrm{\tiny PS}}}U_{\!{\textrm{\tiny HWP}}}\right) \otimes\left( U_{\!{\textrm{\tiny PS}}}U_{\!{\textrm{\tiny BS}}}\right),\nonumber\\  
{\cal U}_3&={\rm CZ}\Big(\left(U_{\!{\textrm{\tiny PS}}}U_{\!{\textrm{\tiny HWP}}}\right)\otimes U_{\!{\textrm{\tiny BS}}}\Big),\nonumber\\  
{\cal U}_4&={\rm CZ}\Big(U_{\!{\textrm{\tiny HWP}}}\otimes\left(U_{\!{\textrm{\tiny PS}}}U_{\!{\textrm{\tiny BS}}}\right)\Big),
\end{align}
where \mbox{$U_{\!{\textrm{\tiny PS}}}={\rm diag}(1,i)$} shifts the phase of the path and polarization qubits by $\pi/2$, and $U_{\!{\textrm{\tiny HWP}}}$ and $U_{\!{\textrm{\tiny BS}}}$ implement the Hadamard gate,
\beq
H=\frac{1}{\sqrt{2}}{\left(\begin{array}{cc}
1&1 \\
1&-1
\end{array}%
\right)},
\eeq
for the polarization qubit and the path qubit, respectively.  For this aim, we use a HWP with its major axis at an angle $\pi/8$ to the optical axis, and a balanced BS.

We see that the unitary transformation ${\cal U}_1$ and ${\cal U}_2$ can be decomposed into a tensor product of two unitary transformations, one for the path qubit and one for the polarization qubit. The unitary transformation ${\cal U}_3$ and ${\cal U}_4$ are not of that kind and could be realized, for example, by using PPBSs together with a Mach-Zehnder interferometer.  This closes our proposal.

To conclude, we are proposing a feasible experimental scheme that implements the SIC POM for a two-qubit system. Our scheme uses linear optical elements and photodetectors, and is, therefore, well within the reach of existing technology. The proposal is based on a successive-measurement approach to SIC POMs. We found that the SIC POM for the qubit pair corresponds to a POM  diagonal in the computational basis, followed by projections onto bases which are mutually unbiased. We observed that this unique construction is owed to a structural relation between the fiducial vectors and the MUB vectors in dimension 4. 

On a more general note, we believe that it would be interesting to learn, if and how this scheme can be generalized to higher dimensions. Such a study could be of a theoretical and a practical use; it might teach us about the SIC POMs' structure in high dimensions and provide new ideas for implementing them. For example, it is possible to show \cite{our} that any SIC POM which is covariant with respect to the Heisenberg-Weyl group can be realized by two successive measurements, each of a rather simple form. This is, in particular, interesting for dimension three, where one has a one-parameter family of non-equivalent group-covariant SIC POMs \cite{appleby05}. 

We would like to thank Huangjun Zhu for valuable and stimulating discussions.
Centre for Quantum Technologies is a Research Centre of Excellence funded by Ministry of Education and National Research Foundation of Singapore.

\end{document}